\documentstyle[12pt]{article}

\newcounter{eq}
\newcounter{sc}

\def\overleftrightarrow#1{\vbox{\ialign{##\crcr
 $\leftrightarrow$\crcr\noalign{\kern-1pt\nointerlineskip}
 $\hfil\displaystyle{#1}\hfil$\crcr}}}

\newlength{\minitwocolumn}
\begin{document}

%%%%%%%%%%%%%%%%%%%%%%%%%%%%%%%%%%%%%%%%%%%%%%%%%%%%%%%%%%%%%%%%%%
%%%%%%%%%%%%%%%%%%%%%%%% Title %%%%%%%%%%%%%%%%%%%%%%%%%%%%%%%%%%%
%%%%%%%%%%%%%%%%%%%%%%%%%%%%%%%%%%%%%%%%%%%%%%%%%%%%%%%%%%%%%%%%%%
\begin{flushright}
DPUR/TH/43\\
March, 2015\\
%hep-th/070****\\
\end{flushright}
\vspace{20pt}

%\magnification=\magstep1
\pagestyle{empty}
\baselineskip15pt
%\font\cmssB=cmss17
%\font\cmssS=cmss10

\begin{center}
{\large\bf Hawking Radiation inside Black Holes in Quantum Gravity
\vskip 1mm }

\vspace{20mm}
Ichiro Oda \footnote{E-mail address:\ ioda@phys.u-ryukyu.ac.jp
}

\vspace{5mm}
           Department of Physics, Faculty of Science, University of the 
           Ryukyus,\\
           Nishihara, Okinawa 903-0213, Japan.\\

\end{center}

%\maketitle

\vspace{5mm}
\begin{abstract}
We study black hole radiation inside black holes within the framework of quantum gravity. 
First, we review on our previous work of a canonical quantization for a spherically symmetric
geometry where one of the spatial coordinates is treated as the time variable, since 
we think of the interior region of a black hole. Based on this formalism, under physically
plausible assumptions, we solve the Wheeler-De Witt equation inside the black hole,
and show that the mass-loss rate of an evaporating black hole due to thermal radiation
is equivalent to the result obtained by Hawking in his semi-classical approach.
A remarkable point is that our assumptions make the momentum constraint coincide with
the Hamiltonian constraint up to an irrelevant overall factor.
Furthermore, for comparison, we solve the Wheeler-De Witt equation outside the black hole
as well, and see that the mass-loss rate of an evaporating black hole has the same expression. 
The present analysis suggests that the black hole radiation comes from the black hole 
singularity. We also comment on the Birkhoff theorem in quantum gravity.
\end{abstract}

\newpage
\pagestyle{plain}
\pagenumbering{arabic}
%\setcounter{page}{1}

%%%%%%%%%%%%%%%%%%%%%%%%%%%%%%%%%%%%%%%%%%%%%%%%%%%%%%%%%%%%%%%%%%
%%%%%%%%%%%%%%%%%%%%%%%% Article %%%%%%%%%%%%%%%%%%%%%%%%%%%%%%%%%
%%%%%%%%%%%%%%%%%%%%%%%%%%%%%%%%%%%%%%%%%%%%%%%%%%%%%%%%%%%%%%%%%%

\rm
%%%%%%%%%%%%%%%%%%%%%%%%%%%%%%%%%%%%%%%%%%%%%%%%%%%%%%%%%%%%%%%%%%%%%
%%%%%%%%%%%%%%%%%%%%%%%%%%%%%%   SEC  1    %%%%%%%%%%%%%%%%%%%%%%%%%%
%%%%%%%%%%%%%%%%%%%%%%%%%%%%%%%%%%%%%%%%%%%%%%%%%%%%%%%%%%%%%%%%%%%%%
\section{Introduction}

There has been a recent revival of an interest in the interior of a black hole. 
For instance, there has been an active debate on whether the AdS/CFT correspondence
could describe the physics of the interior of a black hole or not \cite{Pol1, Pol2, Papa1}. 
This problem is closely related to the information loss paradox in physics of black holes. 
Moreover, it has been more recently pointed out that black holes have a very large interior: 
For a stellar black hole, the volume inside the black hole is larger than that of our 
universe \cite{Chris, Beng}. This study is also relevant to the information loss paradox 
since there might be a lot of real estate available inside a black hole for stocking micro-states 
associated with a black hole entropy. \footnote{See Ref. \cite{Kawai} for a phenomenological 
description of a spherically symmetric black hole.}

In the 1990's, there were also active studies of the interior region of a black hole despite that
the interior is physically of no relevance for external observers outside the horizon.
In those days, the interest in the interior of a black hole has been triggered by
development of the understanding of the internal geometry near the Cauchy horizon inside
the Reissner-Nordstrom black hole, what is called, the mass inflation \cite{Poisson, Ori, Brady}, 
and the phenomenon of the smearing of a black hole singularity in quantum gravity \cite{Hosoya1}. 
As one of motivations behind these studies, there was an expectation that since both the Cauchy horizon 
and a spacetime singularity exhibit highly pathological behavior in the classical theory of general 
relativity, and quantum effects would play a dominant role, studies of the physics inside the horizon 
of a black hole might give us some important clues for constructing a theory of quantum gravity.

Stimulated with the interest in the interior of a black hole in the 1990's, we have already formulated
a canonical formalism of a system with a spherically symmetric black hole holding in the interior region 
bounded by the apparent horizon and the singularity \cite{Hosoya2}, which is a natural generalization of 
the canonical formalism holding in the exterior region covering the spacetime between the apparent horizon 
and the spatial infinity \cite{Hajicek}. In this region, the Killing vector $\frac{\partial}{\partial t}$ 
becomes spacelike whereas it is timelike in the exterior region. Consequently, one has to foliate the interior 
of a black hole with a family of spacelike hypersurfaces, for instance, $r = const$.  

As one of applications of this canonical formalism, following Tomimatsu's idea \cite{Tomimatsu}, we have 
considered black hole radiation in quantum gravity where it was shown that the mass-loss rate 
due to the black hole radiation is equal to that evaluated by Hawking in the semiclassical 
approximation \cite{Hawking}. \footnote{See also related works \cite{Oda1, Hosoya3}.}  In this work, we have 
focused on the vicinity of the apparent horizon where one component $\gamma$ in the metric tensor becomes zero
so we had to adopt a regularization such that $\gamma$ takes a small but finite value. 

This regularization is clearly so unwelcoming that one should dispense with it. In this article, 
without considering the vicinity of the apparent horizon at the beginning, making more physically plausible 
assumptions, we will derive the Hawking radiation in the interior region within the framework of quantum gravity. 
This modification of the model setting makes it possible to calculate the expectation value of the mass-loss rate
without any pathology. Moreover, we will also consider the exterior region of a black hole and do the same job 
in order to make a comparison of black hole radiation between the interior and the exterior of the horizon. 
      
The analysis of both the interior and exterior regions of the apparent horizon of a dynamical
black hole suggests that the thermal radiation of the back hole comes from the spacetime singularity.
This fact might imply that the resolution of the information loss paradox needs the understanding of the
physics of the spacetime singularity of a black hole. As a bonus, we will comment on the Birkhoff 
theorem \cite{Birkhoff} in quantum gravity. 

This article is organized as follows: In the next section, after mentioning notation and conventions,
we review on the canonical formalism of a system with a spherically symmetric black hole in the
interior region bounded by the apparent horizon and the singularity \cite{Hosoya2}.
In Section 3, we apply the canonical formalism reviewed in Section 2 for the calculation of 
the mass-loss rate due to black hole radiation. In Section 4, we also calculate the mass-loss rate
in the exterior region bounded by the apparent horizon and the spatial infinity. The final section contains
a conclusion.

%%%%%%%%%%%%%%%%%%%%%%%%%%%%%%%%%%%%%%%%%%%%%%%%%%%%%%%%%%%%%%%%%%%%%
%%%%%%%%%%%%%%%%%%%%%%%%%%%%%%   SEC  2    %%%%%%%%%%%%%%%%%%%%%%%%%%
%%%%%%%%%%%%%%%%%%%%%%%%%%%%%%%%%%%%%%%%%%%%%%%%%%%%%%%%%%%%%%%%%%%%%
\section{Review of canonical formalism inside black hole}

Before delving into details, let us explain our notation and conventions. We mainly follow 
notation and conventions by Misner et al.'s textbook \cite{MTW}, for instance, the flat Minkowski metric
$\eta_{\mu\nu} = diag(-, +, +, +)$, the Riemann curvature tensor $R^\mu \, _{\nu\alpha\beta} = 
\partial_\alpha \Gamma^\mu_{\nu\beta} - \partial_\beta \Gamma^\mu_{\nu\alpha} + \Gamma^\mu_{\sigma\alpha} 
\Gamma^\sigma_{\nu\beta} - \Gamma^\mu_{\sigma\beta} \Gamma^\sigma_{\nu\alpha}$, and the Ricci tensor 
$R_{\mu\nu} = R^\alpha \, _{\mu\alpha\nu}$. Throughout this article, we adopt the natural units $c = \hbar = G = 1$. 
In this units, all quantities become dimensionless. 

Let us start with the review of a canonical formalism of a spherically symmetric system with a black hole,
which has been presented in our previous work \cite{Hosoya2}, but compared to this work we will change the notation
slightly for the purpose of the comparison with the canonical formalism in the exterior geomery (For instance,
we have exchanged the role between the lapse function $\alpha$ and the shift function $\beta$).
 
As a first step, one needs to select arbitrary spherically symmetric spacelike hypersurfaces to foliate the spacetime.
The key point here is that the radial coordinate plays the role of time in the interior of the horizon in the
spherically symmetric coordinate system. As a simple choice, it is convenient to take the $x^1 = const$ hypersurfaces
to slice the interior region. In the next section, we will choose the simplest case, $x^1 = r$.

The four-dimensional action which we consider in this paper takes the following form:
%**   Action 1 %%%%%%%%%%%%%%%%%%%%%%%%%%%%%%%%%%%%%%%%%%%%%%%%%%%%%%%%%
\begin{eqnarray}
S = \int d^4 x \sqrt{- ^{(4)}g} \left[ \frac{1}{16 \pi} \, ^{(4)}R - \frac{1}{4 \pi} \, ^{(4)}g^{\mu\nu} 
(D_\mu \Phi)^\dagger D_\nu \Phi - \frac{1}{16 \pi} F_{\mu\nu} F^{\mu\nu} \right],
\label{Action 1}
\end{eqnarray}
%%%%%%%%%%%%%%%%%%%%%%%%%%%%%%%%%%%%%%%%%%%%%%%%%%%%%%%%%%%%%%%%%%%
where $\Phi$ is a complex scalar field and its covariant derivative is given by
%**   Cov-derivative %%%%%%%%%%%%%%%%%%%%%%%%%%%%%%%%%%%%%%%%%%%%%%%%%%%%%%%%%
\begin{eqnarray}
D_\mu \Phi = \partial_\mu \Phi + i e A_\mu \Phi,
\label{Cov-derivative}
\end{eqnarray}
%%%%%%%%%%%%%%%%%%%%%%%%%%%%%%%%%%%%%%%%%%%%%%%%%%%%%%%%%%%%%%%%%%%
with $e$ and $A_\mu$ being the electric charge of $\Phi$ and the $U(1)$ gauge field, respectively.
Moreover, as usual, $F_{\mu\nu}$ is the field strength defined as
%**   Field strength %%%%%%%%%%%%%%%%%%%%%%%%%%%%%%%%%%%%%%%%%%%%%%%%%%%%%%%%%
\begin{eqnarray}
F_{\mu\nu} = \partial_\mu A_\nu - \partial_\nu A_\mu.
\label{Field strength}
\end{eqnarray}
%%%%%%%%%%%%%%%%%%%%%%%%%%%%%%%%%%%%%%%%%%%%%%%%%%%%%%%%%%%%%%%%%%%
To clarify the four-dimensional meaning we put the suffix $(4)$ in front of the metric tensor and
the scalar curvature. As a final note, the Greek indices $\mu, \nu, \cdots$ take the four-dimensional
values 0, 1, 2 and 3 whereas the Latin ones $a, b, \cdots$ do the two-dimensional values 0 and 1.
Of course, it is straightforward to include the other matter fields as well as the cosmological constant
in the action (\ref{Action 1}) even if we limit ourselves to the action for simplicity.  

The most general spherically symmetric ansatz for the four-dimensional line element is of form
%**   Line element %%%%%%%%%%%%%%%%%%%%%%%%%%%%%%%%%%%%%%%%%%%%%%%%%%%%%%%%%
\begin{eqnarray}
^{(4)}ds^2 &=& ^{(4)}g_{\mu\nu} dx^\mu dx^\nu, \nonumber\\
&=& g_{ab} dx^a dx^b + \phi^2 ( d\theta^2 + \sin^2 \theta d\varphi^2 ),
\label{Line element}
\end{eqnarray}
%%%%%%%%%%%%%%%%%%%%%%%%%%%%%%%%%%%%%%%%%%%%%%%%%%%%%%%%%%%%%%%%%%%
where the two-dimensional metric $g_{ab}$ and the radial function $\phi$ are the functions of only the 
two-dimensional coordinates $x^a$. The substitution of the ansatz (\ref{Line element}) into the action
(\ref{Action 1}) and then integration over the angular variables $(\theta, \varphi)$ produces the
two-dimensional effective action
%**   2D action  %%%%%%%%%%%%%%%%%%%%%%%%%%%%%%%%%%%%%%%%%%%%%%%%%%%%%%%%%
\begin{eqnarray}
S &=& \frac{1}{2} \int d^2 x \sqrt{-g} \left[ 1 + g^{ab} \partial_a \phi \partial_b \phi 
+ \frac{1}{2} R \phi^2 \right]            \nonumber\\
&-& \int d^2 x \sqrt{-g} \left[ \phi^2 g^{ab} (D_a \Phi)^\dagger D_b \Phi 
+ \frac{1}{4} \phi^2 F_{ab} F^{ab} \right],
\label{2D action}
\end{eqnarray}
%%%%%%%%%%%%%%%%%%%%%%%%%%%%%%%%%%%%%%%%%%%%%%%%%%%%%%%%%%%%%%%%%%%
where we have assumed that $A_a$ and $\Phi$ are also the functions of the two-dimensional coordinates $x^a$
and set $A_\theta = A_\varphi = 0$.

Next let us rewrite the action (\ref{2D action}) in the ADM form. As remarked before, we will regard
the $x^1$ spatial coordinate as time to cover the interior of a black hole by spacelike hypersurfaces.
The appropriate ADM splitting of (1+1)-dimensional spacetime is given by 
%**   ADM %%%%%%%%%%%%%%%%%%%%%%%%%%%%%%%%%%%%%%%%%%%%%%%%%%%%%%%%%
\begin{eqnarray}
g_{ab} = \left(
    \begin{array}{cc}
      \gamma & \beta \\
      \beta & \frac{\beta^2}{\gamma} - \alpha^2
    \end{array}
  \right).
\label{ADM}
\end{eqnarray}
%%%%%%%%%%%%%%%%%%%%%%%%%%%%%%%%%%%%%%%%%%%%%%%%%%%%%%%%%%%%%%%%%%%
The normal unit vector $n^a$ which is orthogonal to the hypersurfaces $x^1 = const$ reads
%**   Normal unit %%%%%%%%%%%%%%%%%%%%%%%%%%%%%%%%%%%%%%%%%%%%%%%%%%%%%%%%%
\begin{eqnarray}
n^a = \left( \frac{\beta}{\alpha \gamma}, \, - \frac{1}{\alpha} \right).
\label{Normal unit}
\end{eqnarray}
%%%%%%%%%%%%%%%%%%%%%%%%%%%%%%%%%%%%%%%%%%%%%%%%%%%%%%%%%%%%%%%%%%%
The induced metric on the hypersurfaces, that is, the projection operator over 
$x^1 = const$ hypersurfaces, is given by
%**   Projection  %%%%%%%%%%%%%%%%%%%%%%%%%%%%%%%%%%%%%%%%%%%%%%%%%%%%%%%%%
\begin{eqnarray}
h^{ab} = g^{ab} + n^a n^b.
\label{Projection}
\end{eqnarray}
%%%%%%%%%%%%%%%%%%%%%%%%%%%%%%%%%%%%%%%%%%%%%%%%%%%%%%%%%%%%%%%%%%%
It is easy to check that $h^{ab}$ is indeed the projection operator by
inserting (\ref{ADM}) and (\ref{Normal unit}) to (\ref{Projection}).  

The extrinsic curvature $K_{ab}$, its trace $K$ and the scalar curvature $R$ are
given by \cite{Wald}
%**   Extrinsic 1 %%%%%%%%%%%%%%%%%%%%%%%%%%%%%%%%%%%%%%%%%%%%%%%%%%%%%%%%%
\begin{eqnarray}
K_{ab} &=& K_{ba} = h_a \, ^c \nabla_c n_b,             \nonumber\\
K &=& g^{ab} K_{ab} = \nabla_a n^a = \frac{1}{\sqrt{-g}} \partial_a ( \sqrt{-g} \, n^a ),
\nonumber\\
R &=& 2 n^a \partial_a K + 2 K^2 - 2 \nabla_c ( n^a \nabla_a n^c ).
\label{Extrinsic 1}
\end{eqnarray}
%%%%%%%%%%%%%%%%%%%%%%%%%%%%%%%%%%%%%%%%%%%%%%%%%%%%%%%%%%%%%%%%%%%
Using Eqs. (\ref{ADM})-(\ref{Extrinsic 1}), a straightforward calculation reveals us
%**   Extrinsic 2 %%%%%%%%%%%%%%%%%%%%%%%%%%%%%%%%%%%%%%%%%%%%%%%%%%%%%%%%%
\begin{eqnarray}
K &=& - \frac{\gamma'}{2 \alpha \gamma} + \frac{\dot{\beta}}{\alpha \gamma}
- \frac{\beta}{2 \alpha \gamma^2} \dot{\gamma},             \nonumber\\
R &=& 2 n^a \partial_a K + 2 K^2 - \frac{2}{\alpha \sqrt{\gamma}} \partial_0 
\left( \frac{\dot{\alpha}}{\sqrt{\gamma}} \right),
\label{Extrinsic 2}
\end{eqnarray}
%%%%%%%%%%%%%%%%%%%%%%%%%%%%%%%%%%%%%%%%%%%%%%%%%%%%%%%%%%%%%%%%%%%
where $\frac{\partial}{\partial x^0} = \partial_0$ and $\frac{\partial}{\partial x^1} = \partial_1$
are also denoted as an overdot and a prime, respectively. With the help of these equations,
one can cast the action (\ref{2D action}) to the form
%**   2D action 2 %%%%%%%%%%%%%%%%%%%%%%%%%%%%%%%%%%%%%%%%%%%%%%%%%%%%%%%%%
\begin{eqnarray}
S &\equiv& \int d^2 x L                  \nonumber\\
&=& \int d^2 x \Biggl[ \frac{1}{2} \alpha \sqrt{\gamma} \Biggl\{ 1 - (n^a \partial_a \phi)^2
+ \frac{1}{\gamma} \dot{\phi}^2 - K n^a \partial_a (\phi^2) 
+ \frac{\dot{\alpha}}{\alpha \gamma} \partial_0 (\phi^2) \Biggr\}     \nonumber\\
&+& \alpha \sqrt{\gamma} \phi^2 \left\{ | n^a D_a \Phi |^2 - \frac{1}{\gamma} | D_0 \Phi |^2 \right\}
+ \frac{1}{2} \alpha \sqrt{\gamma} \phi^2 E^2 \Biggr]        \nonumber\\
&+& \int d^2 x \left[ \frac{1}{2} \partial_a ( \alpha \sqrt{\gamma} K n^a \phi^2 )
- \frac{1}{2} \partial_0 \left( \frac{\dot{\alpha}}{\sqrt{\gamma}} \phi^2 \right) \right],
\label{2D action 2}
\end{eqnarray}
%%%%%%%%%%%%%%%%%%%%%%%%%%%%%%%%%%%%%%%%%%%%%%%%%%%%%%%%%%%%%%%%%%%
where we have defined $E$ as
%**   E  %%%%%%%%%%%%%%%%%%%%%%%%%%%%%%%%%%%%%%%%%%%%%%%%%%%%%%%%%
\begin{eqnarray}
E = \frac{1}{\sqrt{-g}} F_{01} = \frac{1}{\alpha \sqrt{\gamma}} (\dot{A}_1 - A'_0).
\label{E}
\end{eqnarray}
%%%%%%%%%%%%%%%%%%%%%%%%%%%%%%%%%%%%%%%%%%%%%%%%%%%%%%%%%%%%%%%%%%%

Now the differentiation of the action (\ref{2D action 2}) with respect to the spatial derivative of
the canonical variables $\Phi ( \Phi^\dagger ), \phi, \gamma$ and $A_0$ leads to the corresponding 
canonical conjugate momenta $p_\Phi ( p_{\Phi^\dagger} ), p_\phi, p_\gamma$ and $p_A$
%**   Momenta %%%%%%%%%%%%%%%%%%%%%%%%%%%%%%%%%%%%%%%%%%%%%%%%%%%%%%%%%
\begin{eqnarray}
p_\Phi &=& - \sqrt{\gamma} \phi^2 n^a ( D_a \Phi )^\dagger, \quad 
p_\phi = \sqrt{\gamma} n^a \partial_a \phi + \sqrt{\gamma} K \phi,  \nonumber\\
p_\gamma &=& \frac{1}{4 \sqrt{\gamma}}  n^a \partial_a (\phi^2), \quad
p_A = - \phi^2 E.
\label{Momenta}
\end{eqnarray}
%%%%%%%%%%%%%%%%%%%%%%%%%%%%%%%%%%%%%%%%%%%%%%%%%%%%%%%%%%%%%%%%%%%
Then, the Hamiltonian, which is defined as
%**   Hamiltonian 1  %%%%%%%%%%%%%%%%%%%%%%%%%%%%%%%%%%%%%%%%%%%%%%%%%%%%%%%%%
\begin{eqnarray}
H = \int d x^0 \left( p_\Phi \Phi' + p_{\Phi^\dagger} \Phi'^\dagger + p_\phi \phi'
+ p_\gamma \gamma' + p_A A_0' - L \right),
\label{Hamiltonian 1}
\end{eqnarray}
%%%%%%%%%%%%%%%%%%%%%%%%%%%%%%%%%%%%%%%%%%%%%%%%%%%%%%%%%%%%%%%%%%%
is expressed in terms of a linear combination of constraints as expected from diffeomorphism
invariance
%**   Hamiltonian 2  %%%%%%%%%%%%%%%%%%%%%%%%%%%%%%%%%%%%%%%%%%%%%%%%%%%%%%%%%
\begin{eqnarray}
H = \int d x^0 \left( \alpha H_0 + \beta H_1 + A_1 H_2 \right),
\label{Hamiltonian 2}
\end{eqnarray}
%%%%%%%%%%%%%%%%%%%%%%%%%%%%%%%%%%%%%%%%%%%%%%%%%%%%%%%%%%%%%%%%%%%
where $\alpha, \beta$ and $A_1$ are non-dynamical Lagrange multiplier fields, and the Hamiltonian 
constraint, the momentum one and the constraint associated with the $U(1)$ gauge transformation are respectively
given by
%**   Constraints 1 %%%%%%%%%%%%%%%%%%%%%%%%%%%%%%%%%%%%%%%%%%%%%%%%%%%%%%%%%
\begin{eqnarray}
H_0 &=& \frac{1}{\sqrt{\gamma} \phi^2} p_\Phi p_{\Phi^\dagger} - \frac{\sqrt{\gamma}}{2} 
- \frac{\dot{\phi}^2}{2 \sqrt{\gamma}} + \partial_0 \left( \frac{\partial_0 (\phi^2)}{2 \sqrt{\gamma}} \right)
+ \frac{\phi^2}{\sqrt{\gamma}} | D_0 \Phi |^2                                
\nonumber\\
&-& \frac{2 \sqrt{\gamma}}{\phi} p_\phi p_\gamma
+ \frac{2 \gamma \sqrt{\gamma}}{\phi^2} p_\gamma^2 + \frac{\sqrt{\gamma}}{2 \phi^2} p_A^2,
\nonumber\\
H_1 &=& \frac{1}{\gamma} \left[ p_\Phi D_0 \Phi + p_{\Phi^\dagger} (D_0 \Phi)^\dagger \right]
+ \frac{1}{\gamma} p_\phi \dot{\phi} - 2 \dot{p}_\gamma - \frac{1}{\gamma} p_\gamma \dot{\gamma},
\nonumber\\
H_2 &=& - i e \left( p_\Phi \Phi - p_{\Phi^\dagger} \Phi^\dagger \right) - \dot{p}_A. 
\label{Constraints 1}
\end{eqnarray}
%%%%%%%%%%%%%%%%%%%%%%%%%%%%%%%%%%%%%%%%%%%%%%%%%%%%%%%%%%%%%%%%%%%

As is well known, the action can be written as the first-order ADM canonical form by the
dual Legendre transformation 
%**   Legendre  %%%%%%%%%%%%%%%%%%%%%%%%%%%%%%%%%%%%%%%%%%%%%%%%%%%%%%%%%
\begin{eqnarray}
S = \int d x^1 \left[ \int d x^0 \left( p_\Phi \Phi' + p_{\Phi^\dagger} \Phi'^\dagger
+ p_\phi \phi' + p_\gamma \gamma' + p_A A_0' \right) - H \right].
\label{Legendre}
\end{eqnarray}
%%%%%%%%%%%%%%%%%%%%%%%%%%%%%%%%%%%%%%%%%%%%%%%%%%%%%%%%%%%%%%%%%%%
In order to obtain the correct Hamiltonian which yields the Einstein equations through
the Hamilton equations, it is necessary to supplement surface terms to the Hamiltonian
 (\ref{Hamiltonian 2}) \cite{Regge}.
In the formalism at hand, since we take the variation of all the fields to be zero at boundaries,
we do not have to add any surface terms to the Hamiltonian.

%%%%%%%%%%%%%%%%%%%%%%%%%%%%%%%%%%%%%%%%%%%%%%%%%%%%%%%%%%%%%%%%%%%%%
%%%%%%%%%%%%%%%%%%%%%%%%%%%%%%   SEC  3    %%%%%%%%%%%%%%%%%%%%%%%%%%
%%%%%%%%%%%%%%%%%%%%%%%%%%%%%%%%%%%%%%%%%%%%%%%%%%%%%%%%%%%%%%%%%%%%%
\section{Black hole radiation in the interior region}

We are now ready to apply the canonical formalism constructed in the previous section for understanding 
the Hawking radiation \cite{Hawking} from the viewpoint of the internal region of a black hole
in quantum gravity. A similar analysis was performed in our previous work \cite{Hosoya2}
where only the region near the apparent horizon was considered from the outset. In the present article,
we first work with the whole region bounded between the apparent horizon and the spacetime singularity,
make important assumptions on some variables, derive the Wheeler-De Witt equation, and solve it analytically.
After performing this procedure, we impose the condition of the vicinity of the apparent horizon for
compatibility with field equations. We will see that this method nicely overcomes the problem of the
vanishing $\gamma$ variable near the apparent horizon. In the next section, we will apply the same method
to the exterior region bounded between the apparent horizon and the spatial infinity. 

To consider the simplest model of the Hawking radiation, let us switch off the $U(1)$ gauge field and 
treat with the neutral scalar field by which the gauge constraint $H_2$ is identically vanishing and 
the Hamitonian and momentum constraints reduce to the simpler form 
%**   Constraints 2 %%%%%%%%%%%%%%%%%%%%%%%%%%%%%%%%%%%%%%%%%%%%%%%%%%%%%%%%%
\begin{eqnarray}
H_0 &=& - \frac{\sqrt{\gamma}}{2} - \frac{\dot{\phi}^2}{2 \sqrt{\gamma}} 
+ \partial_0 \left( \frac{\partial_0 (\phi^2)}{2 \sqrt{\gamma}} \right)
+ \frac{\phi^2}{\sqrt{\gamma}} ( \partial_0 \Phi )^2                                
- \frac{2 \sqrt{\gamma}}{\phi} p_\phi p_\gamma
+ \frac{2 \gamma \sqrt{\gamma}}{\phi^2} p_\gamma^2,
\nonumber\\
H_1 &=& \frac{1}{\gamma} p_\Phi \partial_0 \Phi
+ \frac{1}{\gamma} p_\phi \dot{\phi} - 2 \dot{p}_\gamma - \frac{1}{\gamma} p_\gamma \dot{\gamma}.
\label{Constraints 2}
\end{eqnarray}
%%%%%%%%%%%%%%%%%%%%%%%%%%%%%%%%%%%%%%%%%%%%%%%%%%%%%%%%%%%%%%%%%%%

Moreover, we will use the ingoing Vaidya metric \cite{Vaidya} to describe the black hole radiation. 
The treatment of the case of the outgoing Vaidya metric can be made in a perfectly analogous manner. 
We therefore define the two-dimensional coordinates $x^a$ as
%**   Advanced time  %%%%%%%%%%%%%%%%%%%%%%%%%%%%%%%%%%%%%%%%%%%%%%%%%%%%%%%%%
\begin{eqnarray}
x^a = ( x^0, x^1 ) = ( v - r, r ),
\label{Advanced time}
\end{eqnarray}
%%%%%%%%%%%%%%%%%%%%%%%%%%%%%%%%%%%%%%%%%%%%%%%%%%%%%%%%%%%%%%%%%%%
where $v$ is the advanced time coordinate. Now let us fix the two-dimensional diffeomorphisms
by the gauge conditions
%**   Gauge %%%%%%%%%%%%%%%%%%%%%%%%%%%%%%%%%%%%%%%%%%%%%%%%%%%%%%%%%
\begin{eqnarray}
g_{ab} = \left(
    \begin{array}{cc}
      \gamma & \beta \\
      \beta & \frac{\beta^2}{\gamma} - \alpha^2
    \end{array}
  \right)
       = \left(
    \begin{array}{cc}
      - \left( 1 - \frac{2M}{r} \right) & \frac{2M}{r} \\
      \frac{2M}{r} & 1 + \frac{2M}{r}
    \end{array}
  \right).
\label{Gauge}
\end{eqnarray}
%%%%%%%%%%%%%%%%%%%%%%%%%%%%%%%%%%%%%%%%%%%%%%%%%%%%%%%%%%%%%%%%%%%
From the gauge conditions (\ref{Gauge}), the two-dimensional line element takes the form
of the Vaidya metric \cite{Vaidya}
%**   2D line element %%%%%%%%%%%%%%%%%%%%%%%%%%%%%%%%%%%%%%%%%%%%%%%%%%%%%%%%%
\begin{eqnarray}
ds^2 &=& g_{ab} dx^a dx^b, \nonumber\\
&=& - \left( 1 - \frac{2M}{r} \right) dv^2 + 2 dv dr.
\label{2D line element}
\end{eqnarray}
%%%%%%%%%%%%%%%%%%%%%%%%%%%%%%%%%%%%%%%%%%%%%%%%%%%%%%%%%%%%%%%%%%%
For a dynamical black hole, we make use of the local definition of the horizon,
namely, the apparent horizon, rather than the global one, the event horizon.
Note that the apparent horizon is now defined as
%**   Apparent horizon %%%%%%%%%%%%%%%%%%%%%%%%%%%%%%%%%%%%%%%%%%%%%%%%%%%%%%%%%
\begin{eqnarray}
x^1 = r = 2 M (x^0, x^1),
\label{Apparent horizon}
\end{eqnarray}
%%%%%%%%%%%%%%%%%%%%%%%%%%%%%%%%%%%%%%%%%%%%%%%%%%%%%%%%%%%%%%%%%%%
where $M = M (x^0, x^1)$ plays the role of the mass function of a black hole.

Since we treat with a massless scalar field which moves along the null geodesics, 
it is natural to assume that the scalar field $\Phi$ depends on only the null coordinate $v$,
by which the mass function $M$ also becomes the function of the advanced time  coordinate
$v$. Furthermore, the radial function $\phi$ is assumed to be a radial coordinate $r$.
Thus, under the situation at hand, we assume  
%**   Assumption %%%%%%%%%%%%%%%%%%%%%%%%%%%%%%%%%%%%%%%%%%%%%%%%%%%%%%%%%
\begin{eqnarray}
\Phi \approx \Phi(v),  \quad M \approx M(v), \quad \phi \approx r.
\label{Assumption}
\end{eqnarray}
%%%%%%%%%%%%%%%%%%%%%%%%%%%%%%%%%%%%%%%%%%%%%%%%%%%%%%%%%%%%%%%%%%% 
Henceforth, we shall use the simbol $\approx$ to indicate the equalities holding under the
assumptions (\ref{Assumption}). It will turn out that these assumptions play a critical role
in simplifying the diffeomorphism constraints and lead to a solvable model of a quantum 
black hole. 

Here it is valuable to check whether the above assumptions (\ref{Assumption}) to be compatible 
with the field equations as follows: The field equations obtained from the action (\ref{2D action})
are given by
%**   Field equations %%%%%%%%%%%%%%%%%%%%%%%%%%%%%%%%%%%%%%%%%%%%%%%%%%%%%%%%%
\begin{eqnarray}
&{}& - \frac{2}{\phi} \nabla_a \nabla_b \phi + \frac{2}{\phi} g_{ab} \nabla_c \nabla^c \phi
+ \frac{1}{\phi^2} g_{ab} \partial_c \phi \partial^c \phi -  \frac{1}{\phi^2} g_{ab}  = 
2 \left( \partial_a \Phi \partial_b \Phi - \frac{1}{2} g_{ab} \partial_c \Phi \partial^c \Phi \right),
\nonumber\\
&{}& \frac{1}{\sqrt{-g}} \partial_a \left( \sqrt{-g} g^{ab} \partial_b \phi \right)
- \frac{1}{2} R \phi = - \phi \partial_a \Phi \partial^a \Phi,
\nonumber\\
&{}& \partial_a \left( \sqrt{-g} \phi^2 g^{ab} \partial_b \Phi \right) = 0.
\label{Field equations}
\end{eqnarray}
%%%%%%%%%%%%%%%%%%%%%%%%%%%%%%%%%%%%%%%%%%%%%%%%%%%%%%%%%%%%%%%%%%%
The Vaidya metric (\ref{2D line element}) gives us the metric tensor
%**   Vaidya %%%%%%%%%%%%%%%%%%%%%%%%%%%%%%%%%%%%%%%%%%%%%%%%%%%%%%%%%
\begin{eqnarray}
g_{ab} = \left(
    \begin{array}{cc}
      g_{vv} & g_{vr} \\
      g_{rv} & g_{rr}
    \end{array}
  \right)
       = \left(
    \begin{array}{cc}
      - \left( 1 - \frac{2M}{r} \right) & 1 \\
      1 & 0
    \end{array}
  \right).
\label{Vaidya}
\end{eqnarray}
%%%%%%%%%%%%%%%%%%%%%%%%%%%%%%%%%%%%%%%%%%%%%%%%%%%%%%%%%%%%%%%%%%%
To check the compatibility of the assumptions (\ref{Assumption}) with the field
equations (\ref{Field equations}), let us make an ansatz that the variables have the form
%**   Ansatz %%%%%%%%%%%%%%%%%%%%%%%%%%%%%%%%%%%%%%%%%%%%%%%%%%%%%%%%%
\begin{eqnarray}
M \approx M(v), \quad \phi \approx r,
\label{Ansatz}
\end{eqnarray}
%%%%%%%%%%%%%%%%%%%%%%%%%%%%%%%%%%%%%%%%%%%%%%%%%%%%%%%%%%%%%%%%%%% 
but the scalar field is still the function of both $v$ and $r$, i.e., $\Phi \approx \Phi(v, r)$.
The first field equation in (\ref{Field equations}) is satisfied if the following relations
hold:
%**   Solution 1 %%%%%%%%%%%%%%%%%%%%%%%%%%%%%%%%%%%%%%%%%%%%%%%%%%%%%%%%%
\begin{eqnarray}
\partial_r \Phi \approx 0, \quad \partial_v \Phi \approx \frac{\sqrt{\partial_v M}}{r}.
\label{Solution 1}
\end{eqnarray}
%%%%%%%%%%%%%%%%%%%%%%%%%%%%%%%%%%%%%%%%%%%%%%%%%%%%%%%%%%%%%%%%%%% 
Then, we find the second equation in (\ref{Field equations}) to be satisfied automatically. Finally, 
using Eq. (\ref{Solution 1}), the third field equation in (\ref{Field equations}) reduces to 
the nontrivial equation
%**   Solution 2 %%%%%%%%%%%%%%%%%%%%%%%%%%%%%%%%%%%%%%%%%%%%%%%%%%%%%%%%%
\begin{eqnarray}
2 r \partial_v \partial_r \Phi + 2 \partial_v \Phi + r \left( 1 - \frac{2 M}{r} \right) 
\partial_r^2 \Phi \approx 0.
\label{Solution 2}
\end{eqnarray}
%%%%%%%%%%%%%%%%%%%%%%%%%%%%%%%%%%%%%%%%%%%%%%%%%%%%%%%%%%%%%%%%%%% 
This equation (\ref{Solution 2}) as well as the latter relation in (\ref{Solution 1}) require us
to limit ourselves to working with the vicinity of the apparent horizon (\ref{Apparent horizon})
\cite{Hosoya2}. In other words, for the consistency of the field equations, in addition to the assumptions
(\ref{Assumption}), one has to supplement one more assumption
%**   Assumption 2 %%%%%%%%%%%%%%%%%%%%%%%%%%%%%%%%%%%%%%%%%%%%%%%%%%%%%%%%%
\begin{eqnarray}
r \approx 2 M(v).
\label{Assumption 2}
\end{eqnarray}
%%%%%%%%%%%%%%%%%%%%%%%%%%%%%%%%%%%%%%%%%%%%%%%%%%%%%%%%%%%%%%%%%%% 
To put differently, with the assumption (\ref{Assumption 2}) we study physics of a black hole near the apparent horizon
inside a black hole. 

After all, given the assumptions (\ref{Assumption}) and (\ref{Assumption 2}), the field equations become 
%**   Solution 3 %%%%%%%%%%%%%%%%%%%%%%%%%%%%%%%%%%%%%%%%%%%%%%%%%%%%%%%%%
\begin{eqnarray}
\partial_r \Phi &\approx& 0,     \nonumber\\
\partial_v \Phi &\approx& \frac{\sqrt{\partial_v M}}{r} \approx \frac{\sqrt{\partial_v M}}{2 M},  \nonumber\\
\partial_r \partial_v \Phi &\approx& \partial_v \partial_r \Phi \approx - \frac{\sqrt{\partial_v M}}{r^2} 
\approx - \frac{\sqrt{\partial_v M}}{4 M^2}.
\label{Solution 3}
\end{eqnarray}
%%%%%%%%%%%%%%%%%%%%%%%%%%%%%%%%%%%%%%%%%%%%%%%%%%%%%%%%%%%%%%%%%%% 
Then, the solution is found to be
%**   Solution 4 %%%%%%%%%%%%%%%%%%%%%%%%%%%%%%%%%%%%%%%%%%%%%%%%%%%%%%%%%
\begin{eqnarray}
\Phi (v, r) = \left( 1 - \frac{2M}{r} \right)^2 \frac{1}{4 \sqrt{\partial_v M}} 
+ \int^v dv \frac{\sqrt{\partial_v M}}{2 M}.
\label{Solution 4}
\end{eqnarray}
%%%%%%%%%%%%%%%%%%%%%%%%%%%%%%%%%%%%%%%%%%%%%%%%%%%%%%%%%%%%%%%%%%% 
As a result, we have
%**   Solution 5 %%%%%%%%%%%%%%%%%%%%%%%%%%%%%%%%%%%%%%%%%%%%%%%%%%%%%%%%%
\begin{eqnarray}
\Phi(v, r) \approx \int^v dv \frac{\sqrt{\partial_v M}}{2 M},
\label{Solution 5}
\end{eqnarray}
%%%%%%%%%%%%%%%%%%%%%%%%%%%%%%%%%%%%%%%%%%%%%%%%%%%%%%%%%%%%%%%%%%% 
which means that one can set $\Phi(v, r) \approx \Phi(v)$ in the vicinity of the apparent horizon.
In this sense, the assumptions (\ref{Assumption}), if the assumption (\ref{Assumption 2}) is added, are at least 
classically consistent with the field equations (\ref{Field equations}).

Next we will turn our attention to quantum theory. In quantum gravity, following Dirac \cite{Dirac}, one must
impose the constraints (\ref{Constraints 2}) on the wave functional $\Psi$ as operator equations to find the
physical state. In general, it is very difficult to solve such constraint equations at the same time.  
In some specific situations, however, the constraints become tractable. For instance, in quantum cosmology, 
the fundamental equation is the Wheeler-De Witt equation, 
$H_0 \Psi = 0$, which is the operator equation associated with the Hamiltonian constraint, since the momentum 
constraint becomes trivial in cosmology owing to the translation invariance of the universe. Of course, solving
the Wheeler-De Witt equation is still a tough work, so some people try to simplify the equation by considering 
minisuperspace.

In this article, our strategy for solving the operator equations $H_0 \Psi = 0, H_1 \Psi = 0$ is similar to
minisuperspace approach in the sense that we set up the assumptions (\ref{Assumption}) to simplify the constraints.
However, we do not impose the assumption (\ref{Assumption 2}) a priori since this assumption is not only unnecessary
in order to simplify the constraint equations but also leads to a problem of the dynamical variable $\gamma$ being zero. 
Indeed, it is remarkable that the assumptions  (\ref{Assumption}) not only make the momentum constraint
agree with the Hamiltonian one up to an overall factor but also reduce the unified operator equation, which is
nothing but the Wheeler-De Witt equation, to be a solvable equation in an analytical manner. 
  
With the two-dimensional coordinates (\ref{Advanced time}), the derivative operators take the form
%**   Derivative %%%%%%%%%%%%%%%%%%%%%%%%%%%%%%%%%%%%%%%%%%%%%%%%%%%%%%%%%
\begin{eqnarray}
\left( \frac{\partial}{\partial x^0},  \frac{\partial}{\partial x^1} \right) 
\equiv \left( \partial_0, \partial_1 \right) = \left( \partial_v, \partial_v + \partial_r \right).
\label{Derivative}
\end{eqnarray}
%%%%%%%%%%%%%%%%%%%%%%%%%%%%%%%%%%%%%%%%%%%%%%%%%%%%%%%%%%%%%%%%%%% 
With the help of Eqs. (\ref{Normal unit}) and (\ref{Gauge}), in case of a real scalar field and vanishing gauge field
the assumptions (\ref{Assumption}) reduce the canonical conjugate momenta (\ref{Momenta}) to the form 
%**   Momenta 2 %%%%%%%%%%%%%%%%%%%%%%%%%%%%%%%%%%%%%%%%%%%%%%%%%%%%%%%%%
\begin{eqnarray}
p_\Phi &\approx& - \phi^2 \partial_v \Phi, \quad 
p_\phi \approx - \frac{1}{\frac{2M}{\phi} - 1} \partial_v M + 1 - \frac{M}{\phi},  \nonumber\\
p_\gamma &\approx& - \frac{\phi}{2} \approx - \frac{r}{2}.
\label{Momenta 2}
\end{eqnarray}
%%%%%%%%%%%%%%%%%%%%%%%%%%%%%%%%%%%%%%%%%%%%%%%%%%%%%%%%%%%%%%%%%%%  
Then, the remarkable point is that the momentum constraint becomes identical with the Hamiltonian constraint up to
an irrelevant overall factor
%**   Constraints 3 %%%%%%%%%%%%%%%%%%%%%%%%%%%%%%%%%%%%%%%%%%%%%%%%%%%%%%%%%
\begin{eqnarray}
\sqrt{\gamma} H_0 &\approx& - \gamma H_1      \nonumber\\
&\approx& \frac{1}{\phi^2} p_\Phi^2 + \left( \frac{2M}{\phi} - 1 \right) 
\left( p_\phi - 1 + \frac{M}{\phi} \right).
\label{Constraints 3}
\end{eqnarray}
%%%%%%%%%%%%%%%%%%%%%%%%%%%%%%%%%%%%%%%%%%%%%%%%%%%%%%%%%%%%%%%%%%%
This compatibility between the momentum and the Hamiltonian constraints justifies the assumptions (\ref{Assumption}) 
in quantum gravity.

The constraint (\ref{Constraints 3}) as an operator equation on the wave functional $\Psi$ gives rise to 
the Wheeler-De Witt equation
%**   WDW 1 %%%%%%%%%%%%%%%%%%%%%%%%%%%%%%%%%%%%%%%%%%%%%%%%%%%%%%%%%
\begin{eqnarray}
\left[ - \frac{1}{\phi^2} \frac{\partial^2}{\partial \Phi^2} + \left( \frac{2M}{\phi} - 1 \right) 
\left( - i \frac{\partial}{\partial \phi} - 1 + \frac{M}{\phi} \right) \right] \Psi = 0.
\label{WDW 1}
\end{eqnarray}
%%%%%%%%%%%%%%%%%%%%%%%%%%%%%%%%%%%%%%%%%%%%%%%%%%%%%%%%%%%%%%%%%%%
It is worthwhile to rewrite this Wheeler-De Witt equation as follows:
%**   WDW 2 %%%%%%%%%%%%%%%%%%%%%%%%%%%%%%%%%%%%%%%%%%%%%%%%%%%%%%%%%
\begin{eqnarray}
i \frac{\partial \Psi}{\partial T} = \left[ p_\Phi^2 - \frac{2M^2}{e^{2MT} + 1} \tanh(MT) \right] \Psi, 
\label{WDW 2}
\end{eqnarray}
%%%%%%%%%%%%%%%%%%%%%%%%%%%%%%%%%%%%%%%%%%%%%%%%%%%%%%%%%%%%%%%%%%%
where we have defined $T = - \frac{1}{2M} \log (\frac{2M}{\phi} - 1)$. This Wheeler-De Witt equation can be
interpreted as the Schrodinger equation with the Hamiltonian $H = p_\Phi^2 - \frac{2M^2}{e^{2MT} + 1} \tanh(MT)$
and the time $T$ in the superspace at hand. It is of interest to note that the superspace time $T$ "stops" on
the apparent horizon owing to gravitational time dilation. On the other hand, the Hamiltonian has a problematic
behavior in that it is not positive semi-definite. 
 
Now it is easy to find a special solution of the Wheeler-De Witt equation (\ref{WDW 1}) by the method of separation
of variables. The result is given by
%**   WDW-solution 1 %%%%%%%%%%%%%%%%%%%%%%%%%%%%%%%%%%%%%%%%%%%%%%%%%%%%%%%%%
\begin{eqnarray}
\Psi = \left( B  e^{\sqrt{A} \Phi} + C  e^{- \sqrt{A} \Phi} \right)  
e^{ i \left[ \phi - M \log \phi - \frac{A}{2M} \log ( \frac{2M}{\phi} - 1 ) \right] }, 
\label{WDW-solution 1}
\end{eqnarray}
%%%%%%%%%%%%%%%%%%%%%%%%%%%%%%%%%%%%%%%%%%%%%%%%%%%%%%%%%%%%%%%%%%%
where $A, B$ and $C$ are integration constants. Provided that the expectation value $< \cal{O} >$ of an operator
$\cal{O}$ is defined as
%**   Exp 1 %%%%%%%%%%%%%%%%%%%%%%%%%%%%%%%%%%%%%%%%%%%%%%%%%%%%%%%%%
\begin{eqnarray}
< {\cal{O}} > = \frac{1}{\int d \Phi |\Psi|^2} \int d \Phi \Psi^* {\cal{O}} \Psi,
\label{Exp 1}
\end{eqnarray}
%%%%%%%%%%%%%%%%%%%%%%%%%%%%%%%%%%%%%%%%%%%%%%%%%%%%%%%%%%%%%%%%%%%
one can calculate the expectation value of mass-loss rate $< \partial_v M >$ by using (\ref{Momenta 2}) or
(\ref{Constraints 3})
%**   Mass-loss 1 %%%%%%%%%%%%%%%%%%%%%%%%%%%%%%%%%%%%%%%%%%%%%%%%%%%%%%%%%
\begin{eqnarray}
< \partial_v M > = - \frac{A}{\phi^2}.  
\label{Mass-loss 1}
\end{eqnarray}
%%%%%%%%%%%%%%%%%%%%%%%%%%%%%%%%%%%%%%%%%%%%%%%%%%%%%%%%%%%%%%%%%%%
At this stage, let us substitute the condition (\ref{Assumption 2}) for the consistency of field equations. 
Then, we obtain
%**   Mass-loss 2 %%%%%%%%%%%%%%%%%%%%%%%%%%%%%%%%%%%%%%%%%%%%%%%%%%%%%%%%%
\begin{eqnarray}
< \partial_v M > = - \frac{k^2}{4 M^2}, 
\label{Mass-loss 2}
\end{eqnarray}
%%%%%%%%%%%%%%%%%%%%%%%%%%%%%%%%%%%%%%%%%%%%%%%%%%%%%%%%%%%%%%%%%%%
where we have set $A = k^2$. This result precisely coincides with that by Hawking in his semiclassical 
approach. Accordingly, our result shows that a black hole completely evaporates within a finite time. 
However, it is worth stressing the difference between the Hawking approach and the present one: In the Hawking
semiclassical approach, the gravitational field is fixed as a classical background, and the matter field
is treated only quantum-mechanically. By contrast, our formulation is purely quantum-mechanical
even if we have imposed physically plausible assumptions on the scalar field, mass function and the radial
field.  

Two remarks are in order. First of all, recall that in our previous article \cite{Hosoya2} we have derived 
the same result for the expectation value of the mass-loss rate (\ref{Mass-loss 2}) in the interior near the
apparent horizon, but we have encountered a difficulty of the dynamical variable $\gamma$ becoming zero, 
by which various equalities become singular. Thus we further had to take a regularization 
such that $\gamma$ is not strictly zero 
but takes a small but finite value. It is worth mentioning that in the present formulation this artificial 
regularization is avoided by imposing the condition (\ref{Assumption 2}) only at the final stage. 

Second, one should comment on the boundary condition on the wave functional $\Psi$.  Our physical state,
which satisfies the Wheeler-De Witt equation, takes the form
%**   WDW-solution 2 %%%%%%%%%%%%%%%%%%%%%%%%%%%%%%%%%%%%%%%%%%%%%%%%%%%%%%%%%
\begin{eqnarray}
\Psi = \left( B  e^{|k| \Phi} + C  e^{- |k| \Phi} \right)  
e^{ i \left[ \phi - M \log \phi - \frac{k^2}{2M} \log ( \frac{2M}{\phi} - 1 ) \right] }.
\label{WDW-solution 2}
\end{eqnarray}
%%%%%%%%%%%%%%%%%%%%%%%%%%%%%%%%%%%%%%%%%%%%%%%%%%%%%%%%%%%%%%%%%%%
This physical state does not satisfy the Dirichlet boundary condition $\Psi \rightarrow 0$ for $|\Phi|
\rightarrow \infty$, nor is its norm $\int d \Phi |\Psi|^2$ finite. 
Of course, these requirements might be too strict since we do not have any
physical principle to pick up the appropriate boundary conditions, and the present formulation does not
provide any information on the correct definition of the inner product. Near the spacetime singularity,
because of the huge quantum effects, the matter field $\Phi$ would fluctuate so strongly that the Dirichlet
boundary condition seems to be appropriate to suppress such a unwieldy behavior of the physical state.

%%%%%%%%%%%%%%%%%%%%%%%%%%%%%%%%%%%%%%%%%%%%%%%%%%%%%%%%%%%%%%%%%%%%%
%%%%%%%%%%%%%%%%%%%%%%%%%%%%%%   SEC  4    %%%%%%%%%%%%%%%%%%%%%%%%%%
%%%%%%%%%%%%%%%%%%%%%%%%%%%%%%%%%%%%%%%%%%%%%%%%%%%%%%%%%%%%%%%%%%%%%
\section{Black hole radiation in the exterior region}

Now we will move on to an application of our idea for the Hawking radiation in the exterior region
of a black hole in quantum gravity. The same problem has been already attacked by Tomimatsu \cite{Tomimatsu}
where the assumption (\ref{Assumption 2}) is fully utilized from scratch. In this section, we will
use only the assumptions (\ref{Assumption}) to find the physical state, and impose the condition (\ref{Assumption 2})
at the final stage in interpreting the mass-loss rate of a dynamical black hole. 

The argument proceeds in a perfectly analogous way to the case of the interior region of a black hole.
In the exterior, the ADM splitting of (1+1)-dimensional spacetime is of form \cite{Hajicek}
%**   ADM 2 %%%%%%%%%%%%%%%%%%%%%%%%%%%%%%%%%%%%%%%%%%%%%%%%%%%%%%%%%
\begin{eqnarray}
g_{ab} = \left(
    \begin{array}{cc}
      \frac{\beta^2}{\gamma} - \alpha^2 & \beta \\
      \beta & \gamma
    \end{array}
  \right).
\label{ADM 2}
\end{eqnarray}
%%%%%%%%%%%%%%%%%%%%%%%%%%%%%%%%%%%%%%%%%%%%%%%%%%%%%%%%%%%%%%%%%%%
The normal unit vector $n^a$ to the Cauchy hypersurfaces $x^0 = const$ reads
%**   Normal unit 2 %%%%%%%%%%%%%%%%%%%%%%%%%%%%%%%%%%%%%%%%%%%%%%%%%%%%%%%%%
\begin{eqnarray}
n^a = \left( \frac{1}{\alpha}, \, - \frac{\beta}{\alpha \gamma} \right).
\label{Normal unit 2}
\end{eqnarray}
%%%%%%%%%%%%%%%%%%%%%%%%%%%%%%%%%%%%%%%%%%%%%%%%%%%%%%%%%%%%%%%%%%%
The trace of the extrinsic curvature is calculated to be
%**   Extrinsic 3 %%%%%%%%%%%%%%%%%%%%%%%%%%%%%%%%%%%%%%%%%%%%%%%%%%%%%%%%%
\begin{eqnarray}
K = \frac{\dot{\gamma}}{2 \alpha \gamma} - \frac{\beta'}{\alpha \gamma}
+ \frac{\beta}{2 \alpha \gamma^2} \gamma'.
\label{Extrinsic 3}
\end{eqnarray}
%%%%%%%%%%%%%%%%%%%%%%%%%%%%%%%%%%%%%%%%%%%%%%%%%%%%%%%%%%%%%%%%%%%

In case of a real scalar field and vanishing gauge field, the canonical conjugate momenta 
$\pi_\Phi, \pi_\phi$, and $\pi_\gamma$ are now given by
%**   Momenta 3 %%%%%%%%%%%%%%%%%%%%%%%%%%%%%%%%%%%%%%%%%%%%%%%%%%%%%%%%%
\begin{eqnarray}
\pi_\Phi &\approx& \phi^2 \partial_v \Phi, \quad 
\pi_\phi \approx \frac{1}{1 + \frac{2M}{\phi}} \partial_v M + \frac{\frac{2M^2}{\phi^2}}{1 + \frac{2M}{\phi}},  
\nonumber\\
\pi_\gamma &\approx& \frac{M}{1 + \frac{2M}{\phi}}.
\label{Momenta 3}
\end{eqnarray}
%%%%%%%%%%%%%%%%%%%%%%%%%%%%%%%%%%%%%%%%%%%%%%%%%%%%%%%%%%%%%%%%%%%  
The momentum constraint turns out to become proportional to the Hamiltonian constraint again
%**   Constraints 4 %%%%%%%%%%%%%%%%%%%%%%%%%%%%%%%%%%%%%%%%%%%%%%%%%%%%%%%%%
\begin{eqnarray}
\sqrt{\gamma} H_0 &\approx& \gamma H_1      \nonumber\\
&\approx& \frac{1}{\phi^2} \pi_\Phi^2 - \left( 1 + \frac{2M}{\phi} \right) \pi_\phi 
+ \frac{2M^2}{\phi^2}.
\label{Constraints 4}
\end{eqnarray}
%%%%%%%%%%%%%%%%%%%%%%%%%%%%%%%%%%%%%%%%%%%%%%%%%%%%%%%%%%%%%%%%%%%
This compatibility between the momentum and the Hamiltonian constraints justifies the assumptions (\ref{Assumption}) 
in quantum gravity as well.

An imposition of the constraint (\ref{Constraints 4}) as an operator equation on the wave functional $\Psi$
produces the Wheeler-De Witt equation
%**   WDW 3 %%%%%%%%%%%%%%%%%%%%%%%%%%%%%%%%%%%%%%%%%%%%%%%%%%%%%%%%%
\begin{eqnarray}
\left[ - \frac{1}{\phi^2} \frac{\partial^2}{\partial \Phi^2} + i \left( 1 + \frac{2M}{\phi} \right) 
\frac{\partial}{\partial \phi} + \frac{2M^2}{\phi^2} \right] \Psi = 0.
\label{WDW 3}
\end{eqnarray}
%%%%%%%%%%%%%%%%%%%%%%%%%%%%%%%%%%%%%%%%%%%%%%%%%%%%%%%%%%%%%%%%%%%
Then, a special solution of the Wheeler-De Witt equation (\ref{WDW 3}) is given by
%**   WDW-solution 3 %%%%%%%%%%%%%%%%%%%%%%%%%%%%%%%%%%%%%%%%%%%%%%%%%%%%%%%%%
\begin{eqnarray}
\Psi = \left( B  e^{\sqrt{A} \Phi} + C  e^{- \sqrt{A} \Phi} \right)  
e^{ i \frac{A - 2M^2}{2M} \log ( 1 + \frac{2M}{\phi} ) }, 
\label{WDW-solution 3}
\end{eqnarray}
%%%%%%%%%%%%%%%%%%%%%%%%%%%%%%%%%%%%%%%%%%%%%%%%%%%%%%%%%%%%%%%%%%%
where $A, B$ and $C$ are integration constants. 
As before, the expectation value of mass-loss rate $< \partial_v M >$ is calculated to be
%**   Mass-loss 3 %%%%%%%%%%%%%%%%%%%%%%%%%%%%%%%%%%%%%%%%%%%%%%%%%%%%%%%%%
\begin{eqnarray}
< \partial_v M > = - \frac{A}{\phi^2}.  
\label{Mass-loss 3}
\end{eqnarray}
%%%%%%%%%%%%%%%%%%%%%%%%%%%%%%%%%%%%%%%%%%%%%%%%%%%%%%%%%%%%%%%%%%%
After the assumption (\ref{Assumption 2}) is inserted to this result, we obtain
%**   Mass-loss 4 %%%%%%%%%%%%%%%%%%%%%%%%%%%%%%%%%%%%%%%%%%%%%%%%%%%%%%%%%
\begin{eqnarray}
< \partial_v M > = - \frac{k^2}{4 M^2}, 
\label{Mass-loss 4}
\end{eqnarray}
%%%%%%%%%%%%%%%%%%%%%%%%%%%%%%%%%%%%%%%%%%%%%%%%%%%%%%%%%%%%%%%%%%%
where we have defined $A = k^2$ again. This result precisely coincides with that obtained
in the interior region of a black hole.

Thus we have shown that the result of the mass-loss rate of a black hole owing to the Hawking
radiation is equivalent between the interior and the exterior of a black hole. This fact 
suggests that the Hawking radiation comes from the spacetime singularity as expected.
Of course, the physical state satisfying the Wheeler-De Witt equation is different between 
the interior and the exterior regions of a black hole, but the reason is connected with the fact
that the Vaidya metric is not the maxmally extended, complete geometry and has a coordinate singularity
at the horizon. 

As a final comment, let us consider the Birkhoff theorem within the present framework. In classical
general relativity, the Birkhoff theorem holds in the spherically symmetric geometry, thereby prohibiting
the existence of the graviton. In this article, we have also considered the spherically symmetric geometry,
so it is natural to ask ourselves if we could get some information on the Birkhoff theorem in quantum 
gravity. The reparametrization invariance in two dimensions allows the dynamical variable $\gamma$ and the 
radial function $\phi$ to remain as physical degrees of freedom except for the matter field $\Phi$. 
As seen in the arguments done thus far, 
the variable $\phi$ is removed via the assumptions (\ref{Assumption}) so it does not play the
role of a dynamical variable. This reduction of the dynamical degree of freedom could be understood
as a result of the momentum constraint. In this context, it is important to notice that 
the condition (\ref{Assumption 2}) is not needed to reduce the momentum constraint.
On the other hand, via the gauge conditions (\ref{Gauge}), 
the variable $\gamma$ is related to the mass function $M(v)$ whose change rate with respect to 
the advanced time $v$ is evaluated in this article.  In this sense, the gravitational mode is left 
in the form of the mass function for a dynamical black hole in quantum gravity even if there is no explicit 
gravitational wave.

%%%%%%%%%%%%%%%%%%%%%%%%%%%%%%%%%%%%%%%%%%%%%%%%%%%%%%%%%%%%%%%%%%%%%
%%%%%%%%%%%%%%%%%%%%%%%%%%%%%%   SEC  5    %%%%%%%%%%%%%%%%%%%%%%%%%%
%%%%%%%%%%%%%%%%%%%%%%%%%%%%%%%%%%%%%%%%%%%%%%%%%%%%%%%%%%%%%%%%%%%%%
\section{Conclusion}

In this article, we have made assumptions (\ref{Assumption}), thereby the momentum constraint becoming
identical with the Hamiltonian one up to an irrelevant overall factor. Moreover, these assumptions
made it possible to solve the Wheeler-De Witt equation in an analytical way. As mentioned before, imposing
these assumptions on the matter field, mass function and the radial field can be understood naturally 
from the physical viewpoint: The massless scalar field propagates along the null geodesics, and the mass
function of a black hole receives influences from such a scalar field, so they are the functions of
only the variable $v$, $\Phi = \Phi(v), M = M(v)$. The radial function plays the role of the radius of a
black hole at the primitive level, so that it is natural to take $\phi = r$.
It seems that these assumptions have more profound mathematical meaning 
rather than mere technical devices. It is known that the advanced time coordinate $v$, i.e., the tortoise
coordinate, makes the $r-t$ plane look "conformal" so that conformal field theory can be applied \cite{Motl}.
Then, our assumptions $\Phi = \Phi(v), M = M(v)$ can be interpreted as the holomorphic (or analytic, or chiral) 
condition, thereby making the complicated constraints associated with two-dimensional diffeomorphisms be tractable 
and solvable analytically.

One of motivations in this paper was to avoid an artificial regularization, which was adopted in our previous
paper \cite{Hosoya2}, such that $\gamma$ takes a small but finite value, and then derive the mass-loss rate of 
a black hole owing to the Hawking radiation in a more reasonable way. Indeed, without this regularization
we have succeeded in deriving the same result as that in \cite{Hosoya2}. In retrospect, the analysis in this article
seems to give the regularization a sound foundation.  

There are a lot of works to be done in future. Firstly, it is interesting to generalize the present formulation to
the Reissner-Nordstrom black hole where we have to pay attention to the constraint associated with the $U(1)$ gauge
transformation. Secondly, it might be possible to relax the assumption $\phi = r$ since this assumption is somewhat
ad hoc in that $\phi$ has a possibility of having a more general function of the $r$ coordinate. 
Thirdly, in a recent progress on large interior of a black hole \cite{Chris}, the most dominant contribution comes
from $r = \frac{3}{2} M$ hypersurface, which is very close to the horizon $r = 2M$. Thus, it would be interesting
to construct a model of quantum black holes holding near $r = \frac{3}{2} M$ rather than the horizon for understanding
this issue in a quantum-mechanical way.   
Finally, it would be valuable to formulate the present formulation in the Kruskal-Szekeres coordinate system since we
can consider both the interior and the exterior regions of a black hole at the same time, and might have
a smoothly interpolating physical state in the both regions. \footnote{The Kruskal-Szekeres coordinate system holds only
for eternal black holes, so some extension of which might be necessary in taking account of a dynamical black hole.} 
We wish to return these problems in future.

%%%%%%%%%%%%%%%%%%%%%%%%%%%%%%%%%%%%%%%%%%%%%%%%%%%%%%%%%%%%%%%%%%
%%%%%%%%%%%%%%%%%%%%%%%% Acknowledgements %%%%%%%%%%%%%%%%%%%%%%%%%%%%%
%%%%%%%%%%%%%%%%%%%%%%%%%%%%%%%%%%%%%%%%%%%%%%%%%%%%%%%%%%%%%%%%%%
\begin{flushleft}
{\bf Acknowledgements}
\end{flushleft}

This work is supported in part by the Grant-in-Aid for Scientific 
Research (C) No. 25400262 from the Japan Ministry of Education, Culture, 
Sports, Science and Technology.

%%%%%%%%%%%%%%%%%%%%%%%% reference %%%%%%%%%%%%%%%%%%%%%%%%%%%%%%%
%%%%%%%%%%%%%%%%%%%%%%%%%%%%%%%%%%%%%%%%%%%%%%%%%%%%%%%%%%%%%%%%%%

\end{document}